\documentclass[superscriptaddress,aps,preprintnumbers,amsmath,amssymb,nofootinbib]{revtex4}
\usepackage{graphicx}
\usepackage{dcolumn}
\usepackage{bm}

\def\slashchar#1{\setbox0=\hbox{$#1$} 
\dimen0=\wd0 
\setbox1=\hbox{/} \dimen1=\wd1 
\ifdim\dimen0>\dimen1 
\rlap{\hbox to \dimen0{\hfil/\hfil}} 
#1 
\else 
\rlap{\hbox to \dimen1{\hfil$#1$\hfil}} 
/ 
\fi}

\def\e{\epsilon}

\def\k{\kappa}

\def\m{\mu}

\def\D{\Delta}

\def\L{\Lambda}

\def\beq{\begin{eqnarray}}
\def\eeq{\end{eqnarray}}

\newcommand{\vev}[1]{ \left\langle {#1} \right\rangle }

\def\tr{\mathop{\rm tr}}

\begin{document}

\preprint{YITP-09-43}
\preprint{SLAC-PUB-13718}

\title{Studying Gaugino Mass in Semi-Direct Gauge Mediation}

\author{M.~Ibe}
\affiliation{%
SLAC National Accelerator Laboratory, Menlo Park, CA 94309, USA
}%
\author{Izawa K.-I.}
\affiliation{%
Yukawa Institute for Theoretical Physics, Kyoto University, Kyoto
606-8502, Japan
}
\affiliation{%
Institute for the Physics and Mathematics of the Universe, University of Tokyo, Chiba 277-8568, Japan}
\author{Y.~Nakai}
\affiliation{%
Yukawa Institute for Theoretical Physics, Kyoto University, Kyoto 606-8502, Japan
}
\begin{abstract}
We study gaugino mass generation in the context of semi-direct gauge mediation models,
where the messengers are charged under both the hidden sector and the standard model gauge groups while they do not play important roles in dynamical supersymmetry breaking. 
We clarify the cancellation of the leading contributions of the supersymmetry breaking effects 
to the gaugino mass in this class of models in terms of the macroscopic effective theory
of the hidden sector dynamics.
We also consider how to retrofit the model so that we obtain the non-vanishing leading contribution
to the gaugino mass.
\end{abstract}
\date{July 17, 2009}
\maketitle

\section{Introduction}

If supersymmetry is hidden in nature above the scale we have explored up to now, 
it helps us to understand the large hierarchy between the 
electroweak scale and the Planck scale. 
It is, however, not straightforward to correctly hide supersymmetry at low energy,
and many years of experience suggest that breaking of the supersymmetry 
must take place in a so-called hidden sector.
So far, a lot of mechanisms have been proposed which communicate the effects of 
supersymmetry breaking in the hidden sector to the supersymmetric standard-model.
Among them, the gauge mediation
\cite{Giu}
is an attractive mechanism since the unwanted flavor-changing neutral processes
are naturally suppressed in this mechanism.

The pivot of gauge mediation consists of messenger fields that are charged under the 
standard model gauge symmetries. 
We classify the models with gauge mediation into two classes;
the one is a class of models where the messenger fields are charged under the hidden gauge 
dynamics and the other is a class of models 
where the messengers are singlet under the hidden dynamics.
In the former class of the models,
we may further divide the models into two-types; 
direct and semi-direct\,\cite{Sei}
(See earlier examples of the models with semi-direct gauge mediation\,\cite{Iza,Ibe}).
Let $S = m + \theta^2 F$ be a representative spurion for  
supersymmetry breaking (presumably with dynamical origin), 
where $\theta$ denotes the superspace coordinate.
The direct gauge mediation is given by a superpotential term $S Q\bar{Q}$ 
with standard model vector-like pairs $Q$ and $\bar Q$ of chiral superfields 
as the mediators with hidden gauge interaction charges
(whose dynamics cause supersymmetry breaking encoded in the $S$ value). 
The semi-direct mediation is given by a superpotential mass term $\m Q \bar Q$ 
with $\m$ a constant and a representative term $S \psi \bar \psi$ 
with a hidden gauge interaction vector-like pair $\psi$ and $\bar\psi$ of
standard-model singlet chiral superfields.

The important differences between the direct and semi-direct gauge mediations
are that the messengers do not play important roles in dynamical supersymmetry breaking
in the semi-direct models.
As a result, in the semi-direct gauge mediation models, the rank of gauge group
in the supersymmetry breaking sector can be smaller than
that in the direct gauge mediation models. 
In this way, we can ameliorate the Landau pole problem in the semi-direct gauge mediation model, which is often encountered in direct gauge mediation models with low-energy supersymmetry breaking.

A difficulty in the semi-direct gauge mediation models, however, is 
a little hierarchy between the gaugino masses and the sfermion masses in 
the supersymmetric standard model sector.
That is,  in the semi-direct models,
the gaugino masses vanish to the leading order
in $F$, while the scalar masses emerge at the leading order in $F$ 
(the so-called gaugino screening \cite{Ark}),
and the gaugino masses are roughly suppressed by $(F/m^2)^2$ compared with
the scalar masses,%
\footnote{Notice that $F/m^2<1$ is required for the messenger sectors not to have
tachyonic modes.}
which leads to a severe "little hierarchy" problem 
without careful tuning between the sizes of $F$ and $m$.

In this paper, we study gaugino mass generation
in strongly coupled semi-direct gauge mediation models
by using the macroscopic effective theory of the hidden strong dynamics
\cite{Ibe}.
As we will show, the gauge coupling constants of the supersymmetric standard model sector
receive non-trivial threshold corrections when we move to the macroscopic
effective theory from the microscopic gauge theory of the hidden gauge dynamics.
Such threshold corrections, which denote the gauge mediation 
effects from the heavy modes in the hidden dynamics, 
play crucial roles to determine the gaugino masses.
We also show how to retrofit the model so that we obtain the non-vanishing leading contribution
to the gaugino mass.

The organization of the paper is as follows.
In section\,\ref{sec:threshold}, we discuss the threshold corrections
to the gauge coupling constants of the supersymmetric standard model
when we move to a macroscopic effective theory of the hidden strong dynamics.
In section\,\ref{sec:screening}, we restate the gaugino screening
in terms of the macroscopic effective theory.
In section\,\ref{sec:retrofit}, we consider a possible retrofit of the 
semi-direct gauge mediation model so that the leading gaugino mass contributions
are not canceled.
The final section is devoted to discussions.

\section{Threshold correction to the spectator gauge coupling}\label{sec:threshold}
In this section, we consider a model with $SU(N_c)\times SU(N_Q)$
gauge interactions, where $SU(N_c)$ is strongly interacting with 
a dynamical scale $\Lambda$ and is identified as the hidden gauge interaction,
while $SU(N_Q)$ is a weakly coupling spectator gauge interaction
whose subgroups are eventually identified 
as the gauge groups of the supersymmetric standard model. 
In the followings, we show that the gauge coupling of the spectator gauge theory 
receives a non-trivial threshold correction when we move to a macroscopic effective theory of the hidden gauge interaction below the dynamical scale $\L$.
As we will show in the next section, the threshold correction plays a crucial role 
to see the gaugino screening in terms of the macroscopic effective theory.

\subsection{Model with $N_f = N_c + 1$}
We begin with a model with $N_\psi$ flavors of the $SU(N_c)$ fundamental
representation $\psi$'s which are 
singlets under the spectator $SU(N_Q)$ gauge group and one flavor of the bi-fundamental
representation $Q$'s of $SU(N_c)\times SU(N_Q)$ (Table\,\ref{tab:micro}).
We choose $N_\psi$ so that the total flavor of the $SU(N_c)$ gauge theory is $N_f = N_\psi + N_Q = N_c + 1$.
\begin{table}[tdp]
\caption{The matter content of the $SU(N_c)\times SU(N_Q)$ model
with $N_f = N_\psi + N_Q = N_c + 1$.
The subgroups of the $SU(N_Q)$ $(N_Q\ge5)$ are identified with the gauge groups 
of the standard model.
The anomalous $U(1)_A$ global symmetry can be treated as a symmetry by
rotating the dynamical scale of $SU(N_c)$, i.e. $\L$, with a charge given in the table.  }
\begin{center}
\begin{tabular}{c|ccccc}\label{tab:micro}
 & $\psi(\times N_\psi)$ &$\bar\psi(\times N_\psi)$ & $Q$ & $\bar{Q}$
 & $\Lambda^{2N_c-1}$
 \\
 \hline
$SU(N_c)$ &  $\mathbf N_c$ & $\mathbf  {\overline{N}}_c$
&  $\mathbf N_c$ & $\mathbf  {\overline{N}}_c$
&{\bf 1}
\\
  $SU(N_Q)$  &  {\bf 1} &  {\bf 1} &  $\mathbf {N}_Q$ & $\mathbf  {\overline{N}}_Q$
  &{\bf 1}
  \\
  $U(1)_R$ & $1-N_c/N_\psi$ &  $1-N_c/N_\psi$ & $1$ & $1$& $0$\\
  $U(1)_A$ & $0$ & $0$& $1$ & $1$& $2N_Q$
\end{tabular}
\end{center}
\end{table}%

At the scale below the dynamical scale $\L$, the dynamics
of the $SU(N_c)$ theory can be described by combining
microscopic matter contents as
\begin{eqnarray}
\label{eq:lowmass}
 M_\psi &=& \psi \bar{\psi}\ ,\cr
 N &=& \bar{\psi}Q\ ,\cr
 \bar{N}&=&\psi \bar{Q}\ ,\cr
 M_Q &=& Q \bar{Q}\ ,\cr
 B_s &=& \epsilon\,\bar{\psi}\cdots\bar{\psi}\bar{Q}\cdots\bar{Q}, 
 \quad (\bar\psi :\times N_\psi-1,\ \bar{Q} :\times N_Q)\ ,\cr
 \bar{B}_s &=& \epsilon\,{\psi}\cdots{\psi}Q\cdots Q\ ,
 \quad (\psi :\times N_\psi-1,\ {Q} :\times N_Q)\ ,\cr
 B_Q &=& \epsilon\,\bar{\psi}\cdots\bar{\psi}\bar{Q}\cdots\bar{Q}, 
 \quad (\bar\psi :\times N_\psi,\ \bar{Q} :\times N_Q-1)\ ,\cr
 \bar{B}_Q &=& \epsilon\,{\psi}\cdots{\psi}Q\cdots Q\ ,
 \quad (\psi :\times N_\psi,\ {Q} :\times N_Q-1)\ ,
\end{eqnarray}
where $\e$ denotes the invariant anti-symmetric tensor of $SU(N_c)$ group
and we have suppressed the indices of the gauge groups and flavors.
The charges of the macroscopic fields under the spectator gauge group 
as well as the relevant global symmetries are given in Table\,\ref{macro}.
The effective superpotential and K\"ahler potential of those macroscopic fields are 
given by\,\cite{Seiberg:1994bz},
\begin{eqnarray}
 W_{eff} &=& \frac{1}{\L^{2N_c-1}} \left( BM\bar{B}- {\rm det}\,M\right)\ ,\cr 
 K_{eff} &=& \L^2 \left( \frac{|M|^2}{\L^4} + \frac{|B|^2}{\L^{2N_c}} + \frac{|\bar {B}|^2}{\L^{2N_c}}
+\cdots \right)\ ,
\label{eq:potentials}
\end{eqnarray}
where 
$M$, $B$, and $\bar{B}$ collectively denote $(M_\psi, N,\bar{N}, M_Q)$, $(B_s,B_Q)$, and $(\bar{B}_s, \bar{B}_Q)$, respectively, and the ellipses denote the higher dimensional operators.
Here, we have omitted $O(1)$ coefficients in the K\"ahler potential terms.
Let us remind ourselves that $M$, $B$ and $\bar{B}$ are massless 
around their origins, which is consistent with the matchings of the anomalies
of symmetries such as $U(1)_R$--$SU(N_Q)^2$ of the microscopic and the macroscopic theories.

\begin{table}[tdp]
\caption{Macroscopic field content of the $SU(N_c)\times SU(N_Q)$ model
at the scale below $\L$.}
\begin{center}
\begin{tabular}{c|cccccccc}\label{macro}
 & $M_\psi$ &$N(\times N_\psi)$ & $\bar{N}(\times N_\psi)$ & $M_Q$
 & $B_s(\times N_\psi)$ & $\bar{B}_s(\times N_\psi)$
  & $B_Q$ & $\bar{B}_Q$
 \\
 \hline
  $SU(N_Q)$  &  {\bf 1} &   $\mathbf {N}_Q$  &  $\mathbf  {\overline{N}}_Q$& $\mathbf {adj + 1}$
  &{\bf 1}   &{\bf 1} &   $\mathbf {N}_Q$  &  $\mathbf  {\overline{N}}_Q$
  \\
  $U(1)_R$ & $2-2N_c/N_\psi$ &  $2-N_c/N_\psi$ & $2-N_c/N_\psi$ & $2$&$N_c/N_\psi$ & $N_c/N_\psi$ & $0$& $0$   \\
  $U(1)_A$ & $0$ & $1$& $1$ & $2$& $N_Q$& $N_Q$& $N_Q-1$ & $N_Q-1$
\end{tabular}
\end{center}
\end{table}%

Let us discuss the anomalies of the classical $U(1)_A$ symmetry 
given in Table\,\ref{tab:micro} against the spectator $SU(N_Q)$ gauge theory 
in both the microscopic and the macroscopic theories.
In the microscopic theory, the $U(1)_A$--$SU(N_Q)^2$ anomaly is given by
\begin{eqnarray}
\label{eq:micro}
 N_c\times 1\ ,
\end{eqnarray}
which comes from the contribution of $Q$ and $\bar Q$ with a $U(1)_A$ charge $1$.
On the contrary, the anomaly in the macroscopic theory is given by
\begin{eqnarray}
\label{eq:macro}
 N_\psi\times 1 + N_Q\times 2 + 1\times (N_Q-1) = N_c + 2 N_Q\ ,
\end{eqnarray} 
where the first contribution comes from $N_\psi$ flavors of $N$ and $\bar N$,
the second one from $M_Q$, and the last one from $B_Q$ and $\bar{B}_Q$.
As a result, we find that the anomalies in both the theories do not match with each other.
There is no surprise in this disagreement because the $U(1)_A$  symmetry
is anomalous to the $SU(N_c)$ gauge theory.

On the other hand, we may make $U(1)_A$ symmetry free of the anomaly
against the $SU(N_c)$ gauge symmetry,  
by rotating the dynamical scale $\L$ along with the $U(1)_A$ symmetry with
a charge given in Table\,\ref{tab:micro}.
Once the classical $U(1)_A$ symmetry is extended in this way, the anomalies in 
both the microscopic and the macroscopic theories must match with each other.
In fact, the anomaly matching is realized by making the following change
to the gauge kinetic function of $SU(N_Q)$:
\begin{eqnarray}
\label{eq:threshold0}
\frac{1}{g_{SU(N_Q)}^2} \to \frac{1}{g_{SU(N_Q)}^2} -\frac{1}{4\pi^2}\log \frac{\L^{2N_c-1}}{M_*^{2N_c-1}}\ ,
\end{eqnarray}
where $M_*$ denotes a scale at which the gauge coupling constant is defined.
With the rotation of $\L$ given in Table\,\ref{tab:micro}, this term contributes to
the anomaly by $-2N_Q$, and hence, by putting this contribution together with 
Eq.\,(\ref{eq:macro}), we reproduce the anomaly in the microscopic theory in Eq.\,(\ref{eq:micro}).%
\footnote{The above additional term corresponds to the ``gaugino counterterm" discussed in Ref.\,\cite{Dine:2007me}.}

The additional term in Eq.\,(\ref{eq:threshold0})
can be interpreted as a threshold correction from the heavy modes of the $SU(N_c)$ gauge theory with masses in a range of $O(\L)$ which do not appear in Eq.\,(\ref{eq:lowmass}).
Notice that the above threshold correction Eq.\,(\ref{eq:threshold0}) 
does not have non-trivial dependence on the fields $M$, $B$, $\bar B$.
If it had field dependences, the function in the logarithm would take zeros at 
some field values. 
This would imply that the $SU(N_c)$ model should possess extra massless modes at such field points, which is quite unlikely.
Therefore, we conclude that the threshold correction to the spectator gauge coupling
is uniquely determined by Eq.\,(\ref{eq:threshold0}).%
\footnote{There can be a numerical factor in front of $\L^{2N_c-1}$ in Eq.\,(\ref{eq:threshold0}),
which can be absorbed by $M_*$ and does not change the following analysis.}

\subsection{Model with $N_f = N_c $}
Next, let us consider to integrate out one of the fundamental representation $\psi$'s.
Here, we introduce masses to $\psi$'s and $Q$'s by,
\begin{eqnarray}
\label{eq:masses}
 W_{\rm tree} = m_i \psi_i\bar{\psi}_i +\m Q\bar{Q}\ ,
\end{eqnarray}
with $m_i,\m \ll \Lambda$.
Then, at the scale below $\L$, the effective potential is again given by,
\begin{eqnarray}
 W_{eff} &=& \frac{1}{\L^{2N_c-1}} \left( BM\bar{B}- {\rm det}\,M\right)
+ m_i M_{\psi_i} + \m M_Q\ .
\label{eq:potentials}
\end{eqnarray}
Notice that even in the presence of the mass terms,
the threshold correction to the spectator gauge coupling 
from the heavy modes in Eq.\,(\ref{eq:threshold0}) is not changed
and does not depend on the masses in Eq.\,(\ref{eq:masses}),
since there are no invariant combinations which show no singularity
in the limit of $m_i, \m \to 0$.

In the meantime, let us assume that $m_1 \gg m_{i>1},\m$ and integrate out the macroscopic fields
which involve $\psi_1$ as constituents
(see the list in Eq.\,(\ref{eq:lowmass})). 
The relevant equation of motion, {\it i.e.}
$F$-term condition in this case,
is that of the ``heavy mode" $M_{\psi_1}$ which  is given by,
\begin{eqnarray}
\label{EQM1}
 \frac{\partial W}{\partial M_{\psi_1}} = 
 -\frac{M_{\psi_2}\cdots M_{\psi_{N_\psi}} M^{N_Q}_Q }{\L^{2N_c-1}}
 + m_1 =0\ .
\end{eqnarray}
Here, we are considering the vacuum around $M_{ij} = M_i \delta_{ij}$ and $B = \bar{B}=0$
and further assuming all the diagonal components of $M_Q$ take the same value.
At this vacuum, the masses of the macroscopic fields which 
involve $\psi_1$ and are charged under $SU(N_Q)$, {\it i.e.} $N_1$ and $B_Q$, are given by%
\footnote{Here, we are using the normalizations of the field given in Eq.\,(\ref{eq:potentials}),
and hence, the masses are not dimension one parameters.}
\begin{eqnarray}
 M_{N_1} &=& \frac{M_{\psi_2}\cdots M_{\psi_{N_\psi}} M_Q^{N_Q-1}}{\L^{2N_c-1}}\ ,\cr
 M_{B_Q} &=& \frac{M_{Q}}{\L^{2N_c-1}}\ .
\end{eqnarray}
After these charged fields are integrated out, the spectator gauge coupling receives
a threshold correction,
\begin{eqnarray}
\left.\frac{1}{{\mit \D} g^2_{SU(N_Q)}}\right|_{N_1,B_Q}
&=&
-\frac{1}{8\pi^2}\log M_{N_1}M_*
-\frac{1}{8\pi^2}\log M_{B_Q}M_*^{2N_c-3}\ , \cr
&=&-\frac{1}{8\pi^2}\log \frac{M_{\psi_2}\cdots M_{\psi_{N_\psi}} M^{N_Q}_Q }{\L^{4N_c-2}M_*^{2-2N_c}}\ ,  \cr
&=&-\frac{1}{8\pi^2}\log \frac{m_1 }{\L^{2N_c-1}M_*^{2-2N_c}}\ ,
\end{eqnarray}
where we have used the equation of motion given in Eq.\,(\ref{EQM1}) in the final expression.

By putting these threshold corrections together with that from the heavy modes 
in Eq.\,(\ref{eq:threshold0}), we obtain a total threshold correction 
to the spectator gauge coupling, 
\begin{eqnarray}
\label{eq:thresholdTOT1}
\left.\frac{1}{{\mit \D} g^2_{SU(N_Q)}}\right|_{N_1,B_Q,{\rm heavy}}
=-\frac{1}{8\pi^2}\log \frac{m_1 \L^{2N_c-1}}{M_*^{2N_c}}
=-\frac{1}{8\pi^2}\log \frac{\L_1^{2N_c}}{M_*^{2N_c}}\ ,
\end{eqnarray}
where we have defined the dynamical scale of the $SU(N_c)$ gauge theory with 
$N_f = (N_\psi-1) + N_Q = N_c$ flavors by
\begin{eqnarray}
 \L_1^{2N_c}  = m_1 \L^{2N_c-1}\ .
\end{eqnarray}

Now, let us remind ourselves that the $N_f = N_\psi +N_Q= N_c+1$   model 
with a decoupled flavor  cannot be distinguished 
from the  $N_f=(N_\psi-1)+N_Q = N_c$ model at the energy scale well below $\L$ and $m_1$. 
Therefore, the threshold correction derived in Eq.\,(\ref{eq:thresholdTOT1}) is nothing 
but the one from the heavy modes in the model with the dynamical scale $\L_1$
and $N_f = N_\psi + N_Q = N_c$ flavors.
In this way, 
we can derive the threshold correction to the spectator gauge coupling from 
the "heavy" modes with masses of the dynamical scale $\L_1$ 
of the  $N_f = N_\psi + N_Q=N_c$  model, which is given by
\begin{eqnarray}
\label{eq:threshold1}
\left.\frac{1}{{\mit \D} g^2_{SU(N_Q)}}\right|_{{\rm heavy},N_f = N_c}
=-\frac{1}{8\pi^2}\log \frac{\L_1^{2N_c}}{M_*^{2N_c}} \ .
\end{eqnarray}

We may check the consistency of the above threshold correction by examining the anomaly
matching of the classical $U(1)_A$ symmetry in the $N_f = N_c$ theory.
The charge assignments of the macroscopic fields and the dynamical scale
are given in Table\,\ref{tab:micro2}.
In the microscopic theory, the $U(1)_A$--$SU(N_Q)^2$ anomaly is given by
\begin{eqnarray}
\label{eq:U1Amicro}
 N_c\times 1\ ,
\end{eqnarray}
as it appeared in the $N_f = N_c + 1$ model.
On the other hand, the anomaly in the macroscopic theory is now given by
\begin{eqnarray}
\label{eq:U1Amacro}
 N_\psi\times 1 + N_Q\times 2 = N_c + N_Q\ .
\end{eqnarray} 
where the first contribution comes from $N_\psi$ flavors of $N$ and $\bar N$,
and the second one from $M_Q$.
Thus, the anomaly matching of the ``quantum" $U(1)_A$ symmetry with a non-trivial
rotation of $\L_1$ is realized only after we add the anomaly contribution ($-N_Q$) 
from the threshold correction from the heavy mode in Eq.\,(\ref{eq:threshold1}).

\begin{table}[tdp]
\caption{The matter content of the $SU(N_c)\times SU(N_Q)$ model
with $N_f = N_\psi + N_Q = N_c $.
}
\begin{minipage}{.48\linewidth}
\begin{center}
\begin{tabular}{c|ccccc}\label{tab:micro2}
 & $\psi(\times N_\psi)$ &$\bar\psi(\times N_\psi)$ & $Q$ & $\bar{Q}$
 & $\Lambda_1^{2Nc}$
 \\
 \hline
$SU(N_c)$ &  $\mathbf N_c$ & $\mathbf  {\overline{N}}_c$
&  $\mathbf N_c$ & $\mathbf  {\overline{N}}_c$
&{\bf 1}
\\
  $SU(N_Q)$  &  {\bf 1} &  {\bf 1} &  $\mathbf {N}_Q$ & $\mathbf  {\overline{N}}_Q$
  &{\bf 1}
  \\
  $U(1)_R$ & $1-N_c/N_\psi$ &  $1-N_c/N_\psi$ & $1$ & $1$& $0$\\
  $U(1)_A$ & $0$ & $0$& $1$ & $1$& $2N_Q$
\end{tabular}
\end{center}
\end{minipage}
\begin{minipage}{.48\linewidth}
\begin{center}
\begin{tabular}{c|cccccccc}\label{tab:macro}
 & $M_\psi$ &$N(\times N_\psi)$ & $\bar{N}(\times N_\psi)$ & $M_Q$
 & $B_s$ & $\bar{B}_s$
 \\
 \hline
  $SU(N_Q)$  &  {\bf 1} &   $\mathbf {N}_Q$  &  $\mathbf  {\overline{N}}_Q$& $\mathbf {adj + 1}$
  &{\bf 1}   &{\bf 1} 
  \\
  $U(1)_R$ & $2-2N_c/N_\psi$ &  $2-N_c/N_\psi$ & $2-N_c/N_\psi$ & $2$&$0$ & $0$
   \\
  $U(1)_A$ & $0$ & $1$& $1$ & $2$& $N_Q$& $N_Q$
\end{tabular}
\end{center}
\end{minipage}
\end{table}%

\subsection{Model with $N_f = N_c-1 $}
We may further integrate out the second flavor, $\psi_2$, by taking $m_2 \gg m_{i>2},\m$, 
and consider the threshold correction in the model with $N_f = N_c - 1$ flavors.
The relevant equation of motion in this case is 
\begin{eqnarray}
\label{eq:EQM1}
 \frac{\partial W}{\partial M_{\psi_2}} = 
 -\frac{M_{\psi_1}M_{\psi_3}\cdots M_{\psi_{N_\psi}} M^{N_Q}_Q }{\L^{2N_c-1}}
 + m_2 =0\ .
\end{eqnarray}
At this vacuum, the macroscopic field $N_2$ decouples with a mass
\begin{eqnarray}
 M_{N_2} =  \frac{M_{\psi_1}M_{\psi_3}\cdots M_{\psi_{N_\psi}} M^{N_Q-1}_Q }{\L^{2N_c-1}}
 = \frac{m_2}{M_Q}\ ,
\end{eqnarray}
and contributes to the threshold correction by
\begin{eqnarray}
\left.\frac{1}{{\mit \D} g^2_{SU(N_Q)}}\right|_{N_2}
=
-\frac{1}{8\pi^2}\log M_{N_2}M_* 
=-\frac{1}{8\pi^2}\log \frac{m_2 M_*}{M_Q}\ .
\end{eqnarray}
By adding this contribution to Eq.\,(\ref{eq:thresholdTOT1}), we obtain
\begin{eqnarray}
\left.\frac{1}{{\mit \D} g^2_{SU(N_Q)}}\right|_{N_1,N_2,B_Q,{\rm heavy}}
=-\frac{1}{8\pi^2}\log \frac{m_2\L_1^{2N_c}}{M_QM_*^{2N_c-1}}\ .
\end{eqnarray}

As a result, by interpreting this threshold correction as the one from the heavy modes
in the model of $SU(N_c)$ gauge theory with $N_f = N_c -1$ flavors, we obtain 
\begin{eqnarray}
\label{eq:threshold2}
\left.\frac{1}{{\mit \D} g^2_{SU(N_Q)}}\right|_{{\rm heavy},N_f=N_c -1}
=-\frac{1}{8\pi^2}\log \frac{\L_2^{2N_c+1}}{\det M_Q^{1/N_Q}M_*^{2N_c-1}}\ ,
\end{eqnarray}
where we have defined the dynamical scale of the model with $N_f = N_c-1 $ by
\begin{eqnarray}
  \L_2^{2N_c +1} = m_2 \L_1^{2N_c}  = m_1 m_2 \L^{2N_c-1}\ .
\end{eqnarray}
In the above expression, we have also taken into account the $SU(N_Q)$ symmetry
by replacing $M_Q $ with $\det M_Q^{1/N_Q}$.

Let us note that the threshold correction from the heavy fields in the
case $N_f = N_c-1$ 
depends on $\det M_Q$ and is singular at $M_Q = 0$.
This singularity is consistent with the fact that the origin of the macroscopic fields 
are removed by the effective potential\,\cite{Affleck:1983mk},
\begin{eqnarray}
W_{\rm ADS}= \frac{\L_2^{2N_c +1}}{\det M}\ ,
\end{eqnarray}
and the anomaly matchings of the global symmetries 
require that the global symmetries are broken spontaneously.
Here, $M$ collectively denotes the meson fields consisting of $N_f = N_c-1$ flavors of $\psi$  and $Q$.

\subsection{Model with $N_f = N_\psi + N_Q $}
By repeating the above discussion, we obtain the threshold correction 
to the spectator gauge coupling 
from the heavy modes in the model with $N_f = N_\psi+N_Q < N_c+1$ flavors
when we move from the microscopic to the macroscopic theory.
The resultant threshold correction is given by
\begin{eqnarray}
\label{eq:thresholdfin}
\left.\frac{1}{{\mit \D} g^2_{SU(N_Q)}}\right|_{{\rm heavy},N_f=N_\psi +N_Q}
=-\frac{1}{8\pi^2}\log \frac{\L_{\rm eff}^{3N_c-N_f}}{\det M_Q^{(N_c-N_f)/N_Q }M_*^{N_c+N_f}}\ ,
\end{eqnarray}
for $N_c + 1 > N_f >1$\ ,%
\footnote{
Here, we have derived this result by deforming the model with $N_f = N_c + 1$ flavors
with mass terms to the constituent fields.
We may also obtain the same result
by applying the anomaly matching condition directly
to the model with $N_f = N_\psi + N_c$ flavors. 
}
while it is given by
\begin{eqnarray}
\label{eq:thresholdfin0}
\left.\frac{1}{{\mit \D} g^2_{SU(N_Q)}}\right|_{{\rm heavy},N_f=N_c + 1}
=-\frac{1}{4\pi^2}\log \frac{\L_{\rm eff}^{2N_c-1}}{M_*^{2N_c-1}}\ ,
\end{eqnarray}
for $N_f = N_c + 1$.
Here, $\L_{\rm eff}$ denotes the dynamical scale of the $SU(N_c)$ gauge theory
with $N_f$ flavors.

Notice that we have not used the equation of motion of $M_Q$ in the above derivations
of the threshold correction.
Thus, the threshold correction derived above are not changed even if we consider 
more generic superpotential of $Q$'s, 
\begin{eqnarray}
 W_{\rm tree} = m_i \psi_i\bar\psi_i + f(Q\bar Q)\ .
\end{eqnarray}
This property of the threshold correction will play an important role to 
discuss a possible retrofit of the semi-direct gauge mediation model
so that the gaugino-screening is overcome.

\section{Gaugino Screening in Macroscopic Effective Theory}\label{sec:screening}
Now, let us discuss the gaugino screening in the semi-direct gauge mediation model.
The gaugino mass at the leading order in $F$
can be extracted from the spurion dependence of the 
gauge coupling constant after the messenger fields are integrated out\,\cite{Giudice:1997ni}, 
{\it i.e.} 
\begin{eqnarray}
\left. m_{\rm gaugino}\right|_{\rm leading} 
= \left.\frac{1}{2} \frac{1}{g_{eff}^2}\right|_{\theta^2} g_{eff}^2\ .
\end{eqnarray}

The gaugino screening in the semi-direct gauge mediation can be understood 
as follows.
In the semi-direct gauge mediation models, the tree-level superpotential is given by
\begin{eqnarray}
 W_{\rm tree}  = S\psi\bar{\psi} + \m Q\bar{Q}\ ,
\end{eqnarray}
where $\psi$ is again a fundamental representation of $SU(N_c)$,
and $Q$'s are bi-fundamental representations of $SU(N_c)\times SU(N_Q)$ 
which play roles of messengers.
The supersymmetry breaking effect is encapsulated in the spurion field
$S = m + \theta^2 F$.%
\footnote{In this paper, we do not address the explicit model of dynamical supersymmetry
breaking in the hidden sector.
Instead, we assume that the supersymmetry breaking effects are effectively encapsulated 
in the spurion $S$.}
In order for the gaugino masses in the supersymmetric standard model 
to be generated at the leading order in $F$ at the one-loop level,
the effective gauge coupling constants must have non-trivial $S$ dependences after the 
messengers are integrated out, such as,
\begin{eqnarray}
 \frac{1}{g_{\rm eff}^2} \sim \frac{c}{8\pi^2} \log \frac{S}{M_*}\ ,
\end{eqnarray}
where $c$ is a numerical coefficient.
Such an $S$ dependence, however, contradicts with the anomaly matching of the $U(1)_R$
symmetry given, for example, in Table\,\ref{tab:micro}
and an appropriate assignment to $S$.
Therefore, the above dependence must be vanishing, and hence,  
the gaugino masses at the leading order in $F$ are vanishing.

In the followings, we reanalyze the gaugino screening in terms of the macroscopic
effective theory. 
The explicit analysis in terms of the macroscopic theory opens new possibilities to extend the semi-direct gauge mediation model with the gaugino mass emerging at the leading order in $F$.

\subsection{Gaugino screening in macroscopic theory}
Here, as an example, we consider the $N_f = N_\psi + N_Q < N_c$ model with a tree level superpotential
\begin{eqnarray}
 W_{\rm tree} = S\psi\bar\psi+\m Q\bar{Q}\ ,
\end{eqnarray}
where $S = m+F\theta^2$ again denotes the spurion and we are assuming $\sqrt{F}\ll m$.
In this model, the effective superpotential of the macroscopic fields
\begin{eqnarray}
 M_\psi = \psi\bar\psi\ ,\quad N=\bar\psi Q\ ,\quad \bar{N} = \psi \bar Q\ ,\quad M_Q = Q\bar{Q}\ ,
\end{eqnarray}
is given by\,\cite{Affleck:1983mk}
\begin{eqnarray}
 W_{\rm eff} =(N_c-N_f)\left(\frac{\L_{\rm eff}^{3N_c-N_f}}{\det M} \right)^{\frac{1}{N_c-N_f}}+SM_\psi +\m M_Q\ .
\end{eqnarray}
Here, again, $M$ denotes the $M_\psi$, $N$, $\bar{N}$, and $M_Q$ collectively.

By using the effective superpotential, we obtain the $F$-term conditions
of the (light) macroscopic fields,
\begin{eqnarray}
 \frac{\partial W}{\partial M_{\psi}} &=& 
 -N_\psi
 \left(\frac{\L_{\rm eff}^{3N_c-N_f}}{\det M} \right)^{\frac{1}{N_c-N_f}}
 \frac{1}{M_\psi}
 + N_\psi S = 0\ ,\cr
  \frac{\partial W}{\partial M_{Q}} &=& 
 -N_Q \left(\frac{\L_{\rm eff}^{3N_c-N_f}}{\det M} \right)^{\frac{1}{N_c-N_f}}
 \frac{1}{M_Q}
 + N_Q \m = 0\ ,
\end{eqnarray}
where we are again considering the vacuum with $M_{ij} = M_i \delta_{ij}$ and 
further assuming all the diagonal components of $M_\psi$ and $M_Q$ take the same values,
respectively.
At this vacuum, the expectation values of $M_Q$ and $M_\psi$ have
non-trivial $F$ dependences,%
\footnote{As long as we are considering the leading order in $F$, we can 
take the $F$-term condition as the relation between the superfields (see appendix\,\ref{sec:adiabatic}). } 
\begin{eqnarray}
\label{eq:MQ}
 M_\psi &=& \left(\frac{\mu^{N_Q} \L_{\rm eff}^{3N_c-N_f} }{S^{N_c-N_\psi}} \right)^{1/N_c}
 \propto
 \left(1-\frac{N_c-N_\psi}{N_c}\frac{F}{m}\theta^2\right)\ ,\cr
 M_Q &=& \left(\frac{S^{N_\psi} \L_{\rm eff}^{3N_c-N_f} }{\m^{N_c-N_Q}} \right)^{1/N_c}
\propto
 \left(1+\frac{N_\psi}{N_c}\frac{F}{m}\theta^2\right)\ .
\end{eqnarray}

Now let us consider the gauge mediation effects when the 
macroscopic fields $N$ and $\bar N$, which are charged under $SU(N_Q)$,
and the adjoint representation in $M_Q$ are integrated out.
The masses of those fields are given by
\begin{eqnarray}
\label{eq:massNM}
 M_N &=& \frac{1}{M_\psi M_Q}\left(\frac{\L_{\rm eff}^{3N_c-N_f}}{\det M} \right)^{\frac{1}{N_c-N_f}}
 = \frac{S}{M_Q}
 \propto \left( 1 + \frac{N_c - N_\psi}{N_c}\frac{F}{m}\theta^2 \right),
 \cr
 M_{M_Q} &=& \frac{1}{M_Q^2}\left(\frac{\L_{\rm eff}^{3N_c-N_f}}{\det M} \right)^{\frac{1}{N_c-N_f}}
  = \frac{\m}{M_Q}
 \propto \left( 1 - \frac{ N_\psi}{N_c}\frac{F}{m}\theta^2 \right).
\end{eqnarray}
Thus, the gaugino mass of the spectator gauge theory obtains a non-trivial contribution 
at the leading order in $F$, which is obtained from the threshold correction,
\begin{eqnarray}
\label{eq:thresholdlight}
\left.\frac{1}{{\mit \D} g^2_{SU(N_Q)}}\right|_{N,M_Q} 
&=&-\frac{N_\psi}{8\pi^2}\log M_NM_*
-\frac{N_Q}{8\pi^2}\log M_{M_Q}M_* \ , \cr
&=&-\frac{1}{8\pi^2}\log \frac{\m^{N_Q} S^{N_\psi} M_*^{N_Q+N_\psi}}{M_Q^{N_Q+N_\psi}}\ , \cr
&\sim&-\frac{1}{8\pi^2}\log \left( 1 + \frac{N_\psi (N_c-N_f)}{N_c}\frac{F}{m}\theta^2 \right)\ .
\end{eqnarray}
From this threshold correction,  we obtain the gaugino mass in the spectator gauge theory; 
\begin{eqnarray}
\label{eq:GMSB1}
\left. m_{\rm gaugino}\right|_{N,M_Q} = -\frac{g_{SU(N_Q)}^2}{16\pi^2}
\frac{N_\psi (N_c-N_f)}{N_c}\frac{F}{m}\ .
\end{eqnarray}

As we have discussed in the previous section, however, we should also pay attention
to the threshold correction from the heavy modes in Eq.\,(\ref{eq:thresholdfin}).
Since it depends on $M_Q$, the threshold correction from the heavy modes also gives
a non-trivial gaugino mass at the leading order in $F$ through the $F$ dependence of the
vacuum expectation value of $M_Q$;
\begin{eqnarray}
\label{eq:thresholdhev}
\left.\frac{1}{{\mit \D} g^2_{SU(N_Q)}}\right|_{{\rm heavy},N_f=N_\psi +N_Q}
&=&-\frac{1}{8\pi^2}\log \frac{\L_{\rm eff}^{3N_c-N_f}}{M_Q^{N_c-N_f}M_*^{N_c+N_f}}\ , \cr
&\sim&-\frac{1}{8\pi^2}\log \left( 1 - \frac{N_\psi (N_c-N_f)}{N_c}\frac{F}{m}\theta^2 \right)\ .
\end{eqnarray}
The resultant contribution to the gaugino mass at the leading order of $F$ from the heavy mode
is given by,
\begin{eqnarray}
\label{eq:GMSBH}
\left. m_{\rm gaugino}\right|_{{\rm heavy},N_f=N_\psi +N_Q} 
= \frac{g_{SU(N_Q)}^2}{16\pi^2}
\frac{N_\psi (N_c-N_f)}{N_c}\frac{F}{m}\ .
\end{eqnarray}
Therefore, we find that both the contributions to the gaugino mass from the light modes 
Eq.\,(\ref{eq:GMSB1}) and the heavy modes Eq.\,(\ref{eq:GMSBH})
are cancelled with each other, which reproduce the gaugino screening.

This cancellation can be seen more concisely in terms of the threshold correction.
By combining the two contributions in Eqs.\,(\ref{eq:thresholdlight})
 and (\ref{eq:thresholdhev}), we obtain the total threshold correction,
\begin{eqnarray}
\label{eq:cancel}
\left.\frac{1}{{\mit \D} g^2_{SU(N_Q)}}\right|_{N,M_Q,{\rm heavy},N_f=N_\psi +N_Q}
&=&-\frac{1}{8\pi^2}\log \frac{\m^{N_Q}S^{N_\psi}\L_{\rm eff}^{3N_c-N_f}}{M_Q^{N_c}M_*^{N_c}}\ ,
\cr
&=&-\frac{1}{8\pi^2}\log \frac{\m^{N_c}}{M_*^{N_c}}\ ,
\end{eqnarray}
where we have used the vacuum expectation value of $M_Q$ in Eq.\,(\ref{eq:MQ}).
From the final expression, we easily see that the threshold correction has
no $S$ dependence, which shows the gaugino screening.

The above arguments show that the vacuum expectation value of $M_Q$ 
plays an important role in the gaugino screening.
This observation sheds light on possibilities that  
we may overcome the gaugino screening by deforming the tree level
superpotential of $Q\bar Q$ so that the total threshold correction depends on $S$.
In fact, we will show that we can make the semi-direct gauge mediation 
have the gaugino mass at the leading order in $F$ by using such a deformation.

\subsection{Semi-direct mediation via ``spurious" dynamical scale}
Before closing this section, we mention another description of the above
semi-direct gauge mediation.
In the above analysis, we have treated $\psi$'s and $Q$'s in a similar way 
as the constituent fields.
We could, however, have integrated out all the $\psi$'s first before moving to the
macroscopic effective theory. 
In this case, the model just looks like the one with $N_f = N_Q$ and the
effective dynamical scale.
\begin{eqnarray}
 {\L_{\rm eff}'}^{3N_c-N_Q} = S^{N_\psi}\L_{\rm eff}^{3N_c-(N_\psi+N_Q)}\ . 
\end{eqnarray}
The important difference from the model in the previous section is that
the effective dynamical scale now plays a role of the spurion of supersymmetry breaking 
which has a non-trivial $\theta^2$ dependence.

In this description, we may again check the gaugino screening by adding the
gauge mediation effect through the effective superpotential
\begin{eqnarray}
 W_{\rm tree} =
 (N_c-N_Q)\left(\frac{{\L_{\rm eff}'}^{3N_c-N_Q}}{\det M_Q} \right)^{\frac{1}{N_c-N_Q}}+ \m M_Q\ ,
\end{eqnarray}
and the one from the heavy modes of the model with $N_f = N_Q$,
\begin{eqnarray}
\left.\frac{1}{{\mit \D} g^2_{SU(N_Q)}}\right|_{{\rm heavy},N_f=N_Q}
=-\frac{1}{8\pi^2}\log \frac{{\L_{\rm eff}'}^{3N_c-N_Q}}{\det M_Q^{(N_c-N_Q)/N_Q }M_*^{N_c+N_Q}}\ .
\end{eqnarray}

\section{Retrofitted Semi-direct Mediation}\label{sec:retrofit}
In the previous section, we discussed the gaugino screening in terms of the effective macroscopic theory.
There, we saw that the vacuum expectation value of $M_Q$ 
plays an important role for the screening. 
In this section, we show that we can retrofit the model of the semi-direct gauge mediation
so that it has the gaugino mass at the leading order in $F$
by deforming the superpotential of $Q\bar Q$. 

As an example of such a deformation, 
let us consider to deform the model with $N_f = N_\psi + N_Q$ by
adding a quartic terms of $Q$'s;
\begin{eqnarray}
 W_{\rm tree} = S\psi\bar\psi + \m\, {\tr}Q\bar{Q}  + \k (\tr Q\bar Q)^2\ ,
\end{eqnarray}
where $\k$ is a numerical coefficient and 
we have written ``$\tr$" explicitly,
which has been suppressed in the preceding sections,
so that the flavor structure of the quartic term is clarified. 
With the above deformation, 
the supersymmetric expressions of the masses of the $N$ and $M_Q$ 
in Eq.\,(\ref{eq:massNM})
are not changed,%
\footnote{If we deform the model by $\tr M_Q M_Q$  instead of $(\tr M_Q)^2$,
the masses of the adjoint component of $M_Q$ are changed.
Our analysis can be extended to a model with such a deformation straightforwardly.} 
while it affects the $F$-term condition of $M_Q$;
\begin{eqnarray}
  \frac{\partial W}{\partial M_{Q}} &=& 
 -N_Q \left(\frac{\L_{\rm eff}^{3N_c-N_f}}{\det M} \right)^{\frac{1}{N_c-N_f}}
 \frac{1}{M_Q}
 + N_Q \m + 2 N_Q \k \tr M_Q = 0\ .
\end{eqnarray}
Therefore, the threshold correction to the $SU(N_Q)$ gauge coupling in the deformed model
is obtained by changing the $\mu$ in Eq.\,(\ref{eq:cancel}) to $\mu+2\k\tr M_Q$, {\it i.e.}
\begin{eqnarray}
\label{eq:thresholddef}
\left.\frac{1}{{\mit \D} g^2_{SU(N_Q)}}\right|_{N,M_Q,{\rm heavy},N_f=N_\psi +N_Q}
&=&-\frac{1}{8\pi^2}\log \frac{(\m+2\,\k\tr M_Q)^{N_c}}{M_*^{N_c}}\ .
\end{eqnarray}
Contrary to the previous result in Eq.\,(\ref{eq:cancel}),
the threshold correction in the deformed model
has a non-trivial $F$ dependence through the vacuum expectation value of $M_Q$.%
\footnote{
Notice that the threshold correction in Eq.\,(\ref{eq:thresholddef}) has
a singularity  at the field value
\begin{eqnarray}
\label{eq:singular}
\m + 2 \k\tr M_Q = 0\ .
\end{eqnarray}
This singularity corresponds to the massless adjoint fields $M_Q$
at this field value.
}

As a result,
we obtain a gaugino mass at the leading order in $F$, 
which is given by
\begin{eqnarray}
\label{eq:gaugino}
 m_{\rm gaugino} = -N_\psi \frac{g_{SU(N_Q)}^2}{8\pi^2}
 \frac{\k F}{\m m} \vev{\tr M_Q}\ ,
 \end{eqnarray}
where we have used
\begin{eqnarray}
\tr M_Q &\simeq& \vev{\tr M_Q}
 \left(1+\frac{N_\psi}{N_c}\frac{F}{m}\theta^2\right)\ ,
 \end{eqnarray}
 with 
 \begin{eqnarray}
\vev{\tr M_Q} &\simeq& N_Q\left(\frac{m^{N_\psi} \L_{\rm eff}^{3N_c-N_f} }{\m^{N_c-N_Q}} \right)^{1/N_c}.
\end{eqnarray}
Here, we have kept only the leading order contribution in $\k$ by assuming $\m > \k \vev{\tr M_Q}$.
Let us remind ourselves that the sfermion masses are also generated at the thresholds at the masses of $N$, $M_Q$  and heavy modes,
and they are roughly given by, 
\begin{eqnarray}
\label{eq:scalar}
 m_{\rm scalar} =\eta \frac{g_{SU(N_Q)}^2 }{16\pi^2} \frac{F}{m},
\end{eqnarray}
where $\eta$ is a numerical coefficient of the order of unity which is non-vanishing
even in the limit of $\k\to 0$.%
\footnote{The coefficient  $\eta$ is non-calculable since it includes the contributions 
from the heavy modes of the strong dynamics.}
Thus, by comparing Eqs.\,(\ref{eq:gaugino}) and (\ref{eq:scalar}), 
we find that the gaugino mass can be comparable to the scalar masses for 
\begin{eqnarray}
 \m \sim \k \vev{\tr M_Q} \sim \k\, \L_{\rm eff}^2
\left(
\frac{\mu}{\L_{\rm eff}}
\right)^{\frac{N_Q}{N_c}-1}
\left(
\frac{m}{\L_{\rm eff}}
\right)^{\frac{N_\psi}{N_c}}\  .
\end{eqnarray}

Finally, we comment on a possible origin of the quartic deformation term $\k (\tr Q\bar{Q})^2$.
From the gaugino mass obtained in Eq.\,(\ref{eq:gaugino}), we need to have
a rather large coupling constant $\k=O(\m/\vev{\tr M_Q})$, so that the gaugino masses are not too 
suppressed compared with the scalar masses.
Such a large coupling can be
realized, for example,  by introducing an extra singlet field $X$ which couples
with $\tr Q\bar Q$ as
\begin{eqnarray}
 W = - \frac{1}{2}m_X X^2 + k\, X \tr Q\bar Q.
\end{eqnarray}
After integrating out $X$, we obtain the desired quartic term with $\k = k^2/(2m_X)$.
Thus, the coupling of order of $\k = O(\m/\vev{\tr M_Q})$ can be realized for 
$m_X = O(\vev{\tr M_Q}/\m)$ and $k = O(1)$.

\section{Discussions}

In this paper, we studied gaugino mass generation in the semi-direct gauge mediation models.
We found that the gaugino mass screening can be understood
as a cancellation of the gaugino mass contributions from the heavy modes and the light modes 
of the hidden gauge dynamics.
In such a cancellation, we showed that the vacuum expectation value of the macroscopic
messenger field plays an important role.
We also proposed how to 
retrofit the semi-direct gauge mediation model so that the gaugino mass emerges 
at the leading order in $F$.

We have not addressed explicit models of dynamical supersymmetry
breaking in the hidden sector.
In order to construct a realistic semi-direct gauge mediation model, we must identify this sector. However, in doing this, we easily encounter one problem for the following reason. That is to say, in general, supersymmetry is restored due to the quartic term in the tree level superpotential even if we take the dynamical model as the supersymmetry breaking sector.%
\footnote{For example, at the singularity in Eq.\,(\ref{eq:singular}),
the effective masses of $Q$'s are vanishing, and hence,
the supersymmetry can be restored around this point.
}
Therefore, it is necessary to find the model whose supersymmetry is
dynamically broken even in the presence of the quartic term. Apparently,
we may consider metastable supersymmetry breaking models, for instance.

It is also interesting to consider the extention of our analysis to the $N_f>N_c+1$ case. One complexity in this case is that we have to estimate the effect of the remaining hidden gauge interaction.

\section*{Acknowledgements}

We would like to thank F.~Yagi for valuable discussion.
M.~I. is grateful to Yukawa Institute for Theoretical Physics for hospitality
where part of this work was done. 
This work was supported by the Grant-in-Aid for Yukawa International
Program for Quark-Hadron Sciences, the Grant-in-Aid
for the Global COE Program "The Next Generation of Physics,
Spun from Universality and Emergence", and
World Premier International Research Center Initiative
(WPI Initiative), MEXT, Japan.
The work of M.~I. was supported by the U.S. Department of Energy under
contract number DE-AC02-76SF00515.

\appendix

\section{Adiabatic supersymmetry breaking}\label{sec:adiabatic}
In this appendix, we consider how the supersymmetry 
breaking effect of the spurion $S = \vev {S}+F_S\theta^2$
affects the $F$-term VEVs of the other superfields (we collectively name them $M$), which are vanishing in the limit 
of $F_S = 0$.
We consider that the superpotential and the K\"ahler potential are given by $W(M,S)$ and $K(M,S)$, respectively.
In this case, the scalar potential is given by
\begin{eqnarray}
 -V &=& K_{M\bar{M}} | F_M |^2 + ( K_{S\bar{M}} F_M^* F_S + c.c.) + K_{S\bar{S}} | F_S |^2
  + (W_M F_M + W_S F_S+c.c.) \cr
  &=&  K_{M\bar{M}} \left| F_M +  \frac{1}{K_{M\bar M}}( {K_{S\bar{M}}}F_S   + W_M^*)  \right|^2 
  + K_{S\bar{S}} | F_S |^2
  -    \frac{1}{K_{M\bar M}}\left|  {K_{S\bar{M}}}F_S   + W_M^*  \right|^2
      + (W_S F_S+c.c.).
  \end{eqnarray}
Thus, we see the minimal of $F_M$ at
\begin{eqnarray}
F_M =-  \frac{1}{K_{M\bar M}}( {K_{S\bar{M}}}F_S   + W_M^*).
\end{eqnarray}
Now, let us assume that the $W_M = 0$ at the vacuum in the limit of $F_S=0$, 
and calculate the $F_M$ at the leading order of $F_S$
by solving $\partial V/\partial M = 0$.
At the leading order of $F_S$, the relevant terms in $\partial V/\partial M$ are
\begin{eqnarray}
\frac{\partial V}{\partial M } &=& \frac{1}{K_{M\bar M}} ( {K_{S\bar{M}}}F_S   + W_M^*) W_{MM} -
W_{MS} F_S \cr
&=&-F_MW_{MM} -
W_{MS} F_S,
\end{eqnarray}
and hence, we obtain the leading contribution to the $F$-term as
\begin{eqnarray}
F_M = -\frac{W_{MS}}{W_{MM}} F_S. \label{ftermnovev}
\end{eqnarray}
Here, all the scalar VEVs in the right hand side are those in the limit of $F_S = 0$.

We compare the result with the supersymmetric VEV which is obtained 
by the  $F$-term condition, {\it i.e.} $W_M =0$.
We assume that the equation of the $F$-term condition 
can be extended to the one between superfields as long as the SUSY
breaking is adiabatic, that is, the SUSY breaking effect of the spurion $S = \vev {S}+F_S\theta^2$ is turned on continuously from the limit of $F_S=0$.
Under this assumption, the $F$-terms in the $F$-term condition satisfy
\begin{eqnarray}
 W_{MM}F_M + W_{SM}F_S = 0.
\end{eqnarray}
By comparing this result with Eq.\,(\ref{ftermnovev}), we find that we can reproduce the leading $F$-term
obtained by using the potential analysis.
Therefore, as long as we are considering the leading effect, we can make a shortcut 
to obtain the $F$-term VEV by using the $F$-term condition.

\section{Threshold correction to spectator $U(1)$ gauge coupling}\label{sec:U(1)}
In section\,\ref{sec:threshold}, we considered the threshold correction 
to the spectator $SU(N_Q)$ gauge coupling from the heavy modes 
in the hidden $SU(N_c)$ gauge dynamics.
In this appendix, we consider the model with a $U(1)$ spectator gauge group
which requires careful attention to discuss the decoupling procedures by
adding mass terms to the $\psi$'s.
For simplicity, we consider the model with $N_Q = 1$ with tree-level superpotential,
\begin{eqnarray}
 W= m_i \psi_i\bar\psi_i + f(Q\bar{Q}),
\end{eqnarray}
where only $Q$'s are charged under the $U(1)$ gauge group with unit charge
while $\psi$'s are singlet.

\begin{table}[tdp]
\caption{The matter content of the $SU(N_c)\times U(1)$ model
with $N_f = N_c +1$.
}
\begin{minipage}{.44\linewidth}
\begin{center}
\begin{tabular}{c|ccccc}\label{tab:microU1}
 & $\psi(\times N_\psi)$ &$\bar\psi(\times N_\psi)$ & $Q$ & $\bar{Q}$
 & $\Lambda^{2Nc-1}$
 \\
 \hline
$SU(N_c)$ &  $\mathbf N_c$ & $\mathbf  {\overline{N}}_c$
&  $\mathbf N_c$ & $\mathbf  {\overline{N}}_c$
&{\bf 1}
\\
  $U(1)$  & $0$ & $0$ & $1$ & $-1$ &$0$
  \\
  $U(1)_R$ & $0$ &  $0$ & $1$ & $1$& $0$\\
  $U(1)_A$ & $0$ & $0$& $1$ & $1$& $2$
\end{tabular}
\end{center}
\end{minipage}
\begin{minipage}{.52\linewidth}
\begin{center}
\begin{tabular}{c|cccccccc}\label{tab:macroU1}
 & $M_\psi$ &$N(\times N_\psi)$ & $\bar{N}(\times N_\psi)$ & $M_Q$
 & $B_s$ & $\bar{B}_s$
  & $B_Q(\times N_\psi)$ & $\bar{B}_Q(\times N_\psi)$
 \\
 \hline
  $U(1)$  &   0 &   $1$  &  $-1$& $0$
  &0   &0 &   $1$  &  $-1$
  \\
  $U(1)_R$ & $0$ &  $1$ & $1$ & $2$&$0$ & $0$ & $1$& $1$   \\
  $U(1)_A$ & $0$ & $1$& $1$ & $2$& $0$& $0$& $1$ & $1$
\end{tabular}
\end{center}
\end{minipage}
\end{table}%

\subsection{Model with $N_f= N_c + 1$}
We begin with the model with $N_f =N_c+1$.
In this model, the effective macroscopic theory can be described by
\begin{eqnarray}
 M_\psi &=& \psi \bar{\psi},\cr
 N &=& \bar{\psi}Q,\cr
 \bar{N}&=&\psi \bar{Q},\cr
 M_Q &=& Q \bar{Q},\cr
 B_Q &=& \epsilon\,{\psi}\cdots{\psi}Q,
  \quad (\psi :\times N_c-1,\  {Q} :\times 1),\cr
\bar{B}_Q &=& \epsilon\,\bar{\psi}\cdots\bar{\psi}\bar{Q},
 \quad (\bar\psi :\times N_c-1,\ \bar{Q} :\times 1),\cr
B_s &=& \epsilon\,{\psi}\cdots{\psi},
  \quad (\psi :\times N_c,\  {Q} :\times 0),\cr
  \bar{B}_s &=& \epsilon\,\bar{\psi}\cdots\bar{\psi},
 \quad (\bar\psi :\times N_c,\  \bar{Q} :\times 0),
\end{eqnarray}
where the definitions of the baryons are different from those in
section\,\ref{sec:threshold}, because the $U(1)$ charges of the baryons are simply the sums of those of their constituents.
The charge assignments of the relevant symmetries are given in Table\,\ref{tab:microU1}.
With these macroscopic fields, the effective potential is again given by Eq.\,(\ref{eq:potentials})
for $W_{\rm tree} = 0$.
The important difference from the $SU(N_Q)$ model is that the threshold correction from the heavy modes.
In this model, the threshold correction to the gauge coupling constant of $U(1)$ from the heavy modes, which satisfy the anomaly matching conditions of the global symmetries in Table\,\ref{tab:microU1}, is given by
\begin{eqnarray}
 \left.\frac{1}{{\mit \D} g^2_{U(1)}}\right|_{{\rm heavy}, N_f = N_c+1}
&=&
-\frac{N_c}{8\pi^2}\log \frac{\L^{2N_c-1}}{M_*^{2N_c-1}}.
\end{eqnarray}

\subsection{Model with $N_f= N_c $}
In order to obtain the threshold correction to the $U(1)$ gauge coupling in
the $N_f = N_c$ model,
let us consider to make $\psi_1$ heavy by switching on the mass term in $W_{\rm tree}$.
Notice again that the threshold correction from the heavy modes does not change even in the presence of the mass term of $\psi_1$.
In this case, the relevant equation of motion
of the heavy mode, $M_{\psi1}$, is given in Eq.\,(\ref{EQM1}),
by assuming that only the diagonal components obtain non-vanishing values.
Around this point, the heavy fields which are charged under $U(1)$ are $N_1 = \psi_1 Q$, $\bar{N}_1$, $B_{Qi} = \e \psi_1\cdots \psi_{i-1} \psi_{i+1}\cdots Q$, and $\bar{B}_{Qi}$, and they decouple at 
 \begin{eqnarray}
M_{N_1} &=& \frac{M_{\psi_2}\cdots M_{\psi_{N_\psi}} }{\L^{2N_c-1}},\cr
M_{B_{Qi}} &=& \frac{M_{\psi_i}}{\L^{2N_c-1}},\quad (i=2,\cdots,N_c).
\end{eqnarray}
Hence, the threshold corrections from these modes are given by
\begin{eqnarray}
  \left.\frac{1}{{\mit \D} g^2_{U(1)}}\right|_{N_1,B_{Qi}(i=2\cdots N_c)}
&=&
-\frac{1}{8\pi^2}\log {M_{N_1}M_*} 
- \frac{1}{8\pi^2}\log M_{B_{Q2}}\cdots M_{B_{QN_c}} (M_*^{2N_c-3})^{N_c-1} \cr
&=&-\frac{1}{8\pi^2}\log \frac{ (M_{\psi_2}\cdots M_{\psi_{N_c}})^2} 
{\L^{N_c(2N_c-1)}M_*^{-2N_c^2+5N_c-4} } \cr
&=&-\frac{1}{8\pi^2}\log \frac{ m_1^2} 
{M_Q^2\L^{(N_c-2)(2N_c-1)}M_*^{-2N_c^2+5N_c-4} },
\end{eqnarray}
where we have used the equation of motion Eq.\,(\ref{EQM1}),
$(M_{\psi_2}\cdots M_{\psi_{N_c}}) = m_1 \L^{2N_c-1}/M_Q$, in the final expression.

Therefore, putting the heavy modes and $N,B_Q$ contributions together, we obtain 
the total threshold correction which can be identified with the
correction to the $N_f = N_c$ model:
\begin{eqnarray}
  \left.\frac{1}{{\mit \D} g^2_{U(1)}}\right|_{{\rm heavy}, N_f = N_c}=
  \left.\frac{1}{{\mit \D} g^2_{U(1)}}\right|_{N_1,B_{Qi}(i=2\cdots N_c),{\rm heavy},N_f = N_c + 1}
&=&-\frac{1}{8\pi^2}\log \frac{ (m_1\L^{2N_c-1})^2 } 
{M_Q^2M_*^{4N_c-4} } \cr
&=&-\frac{1}{4\pi^2}\log \frac{ \L_1^{2N_c} } 
{M_QM_*^{2N_c-2} },
\end{eqnarray}
where we have defined the dynamical scale of the $N_f = N_c$ model by $\L_1^{2N_c} = m_1 \L^{2N_c-1}$.

\subsection{Model with $N_f = N_c-1$}
Now let us move on to the model with $N_f = N_c -1$ by integrating out $\psi_2$ with a mass $m_2$.
In this case, $N_2$ and $B_{Q1}$ decouple in addition to $N_1$ and $B_{Qi}$ discussed above at
\begin{eqnarray}
M_{N_2} &=& \frac{M_{\psi_1}M_{\psi_3}\cdots M_{\psi_{N_\psi}} }{\L^{2N_c-1}}
= \frac{m_2}{M_Q},\cr
M_{B_{Q1}} &=& \frac{M_{\psi_1}}{\L^{2N_c-1}} = \frac{m_2}{M_Q\det M_{N_c-2}},
\end{eqnarray}
where $M_{N_c -2}$ denotes the mesons that consist of $\bar \psi_i \psi_j$, ($i,j=3,\cdots,N_c$).
Here, we have used the equation of motion of the heavy mode $M_{\psi_2}$,
\begin{eqnarray}
 \frac{\partial W}{\partial M_{\psi_2}} &=& -\frac { M_{\psi_1} M_{\psi_3}\cdots M_{\psi_{N_c}} M_Q}{\L^{2N_c-1}} + m_2 = 0.
\end{eqnarray}

As a result of the decoupling of $M_{N_2}$ and $M_{B_{Q1}}$, we obtain 
the threshold correction as
\begin{eqnarray}
  \left.\frac{1}{{\mit \D} g^2_{U(1)}}\right|_{N_2,B_{Q1}}
&=&
-\frac{1}{8\pi^2}\log {M_{N_2}M_*} 
- \frac{1}{8\pi^2}\log M_{B_{Q1}} M_*^{2N_c-3} \cr
&=&-\frac{1}{8\pi^2}\log \frac{m_2^2 M_*^{2N_c-2}}{M_Q^2 \det M_{N_c-2}}.
\end{eqnarray}
Thus, we obtain the total threshold correction from the heavy modes 
in the $N_f = N_c-1$ model,
\begin{eqnarray}
  \left.\frac{1}{{\mit \D} g^2_{U(1)}}\right|_{{\rm heavy}, N_f = N_c-1}=
  \left.\frac{1}{{\mit \D} g^2_{U(1)}}\right|_{N_2,B_{Q1},{\rm heavy},N_f = N_c}
&=&-\frac{1}{8\pi^2}\log \frac{ m_2^2\L_1^{4N_c}  } 
{M_Q^4\det M_{N_c-2}M_*^{2N_c-2} } \cr
&=&-\frac{1}{8\pi^2}\log \frac{ \L_2^{4N_c+2}  } 
{M_Q^4\det M_{N_c-2}M_*^{2N_c-2} } \cr
&=&-\frac{1}{8\pi^2}\log 
\left(
\frac{ \L_2^{2N_c+1}  } 
{M_Q^3 M_*^{2N_c-5} }
\right)
\left(\frac{ W_{{\rm ADS}, N_c-1} } 
{M_*^{3} }\right).
\end{eqnarray}
Here, we have defined the dynamical scale of the $N_f = N_c-1$ model by
$\L_2^{2N_c+1} = m_2 \L_1^{2N_c}$, and denoted the Affleck-Dine-Seiberg
superpotential\,\cite{Affleck:1983mk}
 that consists of $M_Q$ and
the remaining $M_\psi$s as
\begin{eqnarray}
 W_{{\rm ADS}, N_c-1} = \frac{\L_2^{2N_c+1}}{M_Q\det M_{N_c-2}}.
\end{eqnarray}

\subsection{Model with $N_f= N_\psi + 1$}
By repeating the above procedure, we end up with the threshold correction
\begin{eqnarray}
\label{eq:thresholdfinU1}
\left.\frac{1}{{\mit \D} g^2_{U(1)}}\right|_{{\rm heavy},N_f=N_\psi +N_Q}
=-\frac{1}{8\pi^2}\log \left(\frac{\L_{\rm eff}^{3N_c-N_f}}{ M_Q^{N_c-N_f+2}M_*^{N_c+N_f-4}}\right)
\left(\frac{ W_{{\rm ADS}, N_f} } 
{M_*^{3} }\right),
\end{eqnarray}
with 
\begin{eqnarray}
 W_{{\rm ADS}, N_f} = (N_c-N_f)\left(\frac{\L_{\rm eff}^{3N_c-N_f}}{M_Q\det M_{N_f-1}}\right)^{\frac{1}{N_c-N_f}},
\end{eqnarray}
for $N_c  > N_f >1$,
while it is given by
 \begin{eqnarray}
\label{eq:thresholdfin0U1}
\left.\frac{1}{{\mit \D} g^2_{U(1)}}\right|_{{\rm heavy},N_f=N_c}
=-\frac{1}{4\pi^2}\log \frac{\L_{\rm eff}^{2N_c}}{M_QM_*^{2N_c-2}},
\end{eqnarray}
for $N_f = N_c$ and
\begin{eqnarray}
\label{eq:thresholdfin0U1}
\left.\frac{1}{{\mit \D} g^2_{U(1)}}\right|_{{\rm heavy},N_f=N_c + 1}
=-\frac{N_c}{8\pi^2}\log \frac{\L_{\rm eff}^{2N_c-1}}{M_*^{2N_c-1}},
\end{eqnarray}
for $N_f = N_c + 1$.
Here, $\L_{\rm eff}$ denotes the dynamical scale of the $SU(N_c)$ gauge theory
with $N_f$ flavors.
By means of the above threshold corrections, we can check the gaugino screening 
for $f(M_Q) = \m M_Q$.

\end{document}